\documentclass[twocolumn]{article}
\usepackage[utf8]{inputenc}
\usepackage[T1]{fontenc}
\usepackage[english]{babel}
\usepackage[natbibapa]{apacite}
\usepackage{ifpdf,amsmath,amsthm,amssymb,amsfonts,newtxtext,newtxmath} 
\usepackage{array,graphicx,dcolumn,multirow,hevea,abstract,hanging}
\usepackage[labelfont=sc,textfont=sf]{caption}
\usepackage[hyperfootnotes=false,breaklinks=true]{hyperref} 
\usepackage[hyphenbreaks]{breakurl}
\usepackage{booktabs} 
\newcolumntype{L}[1]{>{\raggedright\arraybackslash }p{#1}} 
\newcolumntype{C}[1]{>{\centering\arraybackslash }p{#1}}
\newcolumntype{R}[1]{>{\raggedleft\arraybackslash }p{#1}}
\newcolumntype{d}[1]{D{.}{.}{#1}} 
\newcommand{\mc}{\multicolumn}
\let\tempone\itemize
\let\temptwo\enditemize
\let\tempthree\enumerate
\let\tempfour\endenumerate
\renewenvironment{itemize}{\tempone\setlength{\itemsep}{0pt}}{\temptwo}
\renewenvironment{enumerate}{\tempthree\setlength{\itemsep}{0pt}}{\tempfour}

\title{``Do the Right Thing'' for Whom? An Experiment on Ingroup Favouritism, Group Assorting and Moral Suasion}

\author{
Ennio Bilancini\thanks{IMT School of Advanced Studies, Lucca. Email: ennio.bilancini@imtlucca.it.}
\and 
  Leonardo Boncinelli\thanks{Department of Economics and Management, University of
  Florence. Email: leonardo.boncinelli@unifi.it.} 
\and 
  Valerio Capraro\thanks{Department of Economics, Middlesex University London. Email: v.capraro@mdx.ac.uk.}
\and 
  Tatiana Celadin\thanks{IMT School of Advanced Studies, Lucca. Email: tatiana.celadin@imtlucca.it.}
\and
  Roberto Di Paolo\thanks{Department of Economics, University of Alicante; IMT School of Advanced Studies, Lucca. Email: roberto.dipaolo@imtlucca.it .} 
}

\date{} 
\begin{document} 

\newcommand{\jref}{http://journal.sjdm.org/vol13.1.html}
\newcommand{\jhead}{Judgment and Decision Making}
\newcommand{\jdate}{Forthcoming}
\pagestyle{myheadings} \markright{\protect\small \href{\jref}{\jhead}, \jdate \hfill ``Do the Right Thing'' for Whom? \qquad}
\begin{htmlonly}
\href{\jref}{\jhead}, \jdate, pp.\
\end{htmlonly}
\twocolumn[
\vspace{-.3in}
{\small \href{\jref}{\jhead}, \jdate
}

\maketitle

\vspace{-3mm}
\begin{onecolabstract}
In this paper we investigate the effect of moral suasion on ingroup favouritism. We report a well-powered, pre-registered, two-stage 2x2 mixed-design experiment. In the first stage, groups are formed on the basis of how participants answer to a set of questions, concerning non-morally relevant issues in one treatment (assorting on non-moral preferences), and morally relevant issues in another treatment (assorting on moral preferences). In the second stage, participants choose how to split a given amount of money between participants of their own group and participants of the other group, first in the baseline setting and then in a setting where they are told to do what they believe to be morally right (moral suasion). Our main results are: (i) in the baseline, participants tend to favour their own group to a greater extent when groups are assorted according to moral preferences, compared to when they are assorted according to non-moral preferences; (ii) the net effect of moral suasion is to decrease ingroup favouritism, but there is also a non-negligible proportion of participants for whom moral suasion increases ingroup favouritism; (iii) the effect of moral suasion is substantially stable across group assorting and four pre-registered individual characteristics (gender, political orientation, religiosity, pro-life vs pro-choice ethical convictions).

\end{onecolabstract}\bigskip
]




\saythanks
\section{Introduction}

In the past years, behavioural scientists have provided converging evidence that social behaviour in economic games is not solely motivated by the monetary consequences of the available actions, but it is also motivated by moral preferences for doing what people think to be the right thing, beyond the monetary consequences that this action brings about. For example, Krupka and Weber (\citeyear{krupka2013identifying}) found that people tend to be more altruistic in the dictator game in the ``take frame'', compared to the dictator game in the ``give frame'', in spite of the fact that these two frames give rise to monetarily equivalent decision problems; and, crucially, they showed that this framing effect can be explained by preferences for making the socially appropriate choice: people tend to rate ``taking from the recipient'' to be more socially inappropriate than ``not giving to the recipient''. A conceptually similar result has been obtained for six dictator game frames in Capraro and Vanzo (\citeyear{capraro2019power}), where the authors found that their framing effects can be explained by a change in the perception of what people think is the morally right thing to do. The fact that frames can impact people's choices by activating moral preferences is not limited to altruistic behaviour in the dictator game. For example, Eriksson \emph{et al}. (\citeyear{eriksson2017costly}) reported that moral considerations explain framing effects in ultimatum game rejections, whereas Capraro and Rand (\citeyear{capraro2018right}) demonstrated that moral preferences explain framing effects in trade-off games pitting equity against efficiency, and drive not only altruistic behaviour in the dictator game, but also cooperative behaviour in the prisoner’s dilemma. Besides the empirical work, moral preferences have also been formalized in several economic models \citep{benabou2006incentives,levitt2007laboratory,lopez2008aversion,andreoni2009social,dellavigna2012testing,kessler2012norms,alger2013homo,krupka2013identifying,kimbrough2016norms}. And besides behavioural economics, the fact that at least altruistic giving is driven by morality has been highlighted by several scholars, to the point that fairness is considered to be one of foundations of morality \citep{haidt2004intuitive,graham2013moral,haidt2012righteous}.

One of the main applications of this literature on moral preferences is the work on moral suasion. The idea is simple: if people’s behaviour is driven by moral preferences, then making morality salient should impact people’s behaviour. A number of works have provided support for this hypothesis. An earlier paper by Bra{\~n}as-Garza (\citeyear{branas2007promoting}) found that telling dictators that ``the other person relies on you'' increases dictator game donations towards that person. A subsequent work by Dal B\'o and Dal B\'o (\citeyear{dal2014right}) reported that reminders of the Golden Rule increases cooperation in an iterated prisoner’s dilemma. More recently, Capraro \emph{et al}. (\citeyear{capraro2019increasing}) showed that asking participants ``What do you personally think is the morally the right thing to do?'', prior to making their decision, increases dictator game altruism and prisoner’s dilemma cooperation. Scholars have also started applying moral suasion to redistribution decisions that have consequences outside the laboratory. Capraro \emph{et al}. (\citeyear{capraro2019increasing}) found that the aforementioned moral nudge increases online charitable donations to humanitarian organizations by 44\%. Bott \emph{et al}. (\citeyear{bott2019you}) reported that sending moral letters to tax payers decreases tax evasion. In sum, moral suasion is emerging as a useful tool to nudge prosocial behaviour both in and outside the lab.\footnote{A related stream of literature uses descriptive and injunctive norms to promote prosocial behaviour in the laboratory \citep{bicchieri2009right,krupka2009focusing,zafar2011experimental,raihani2014dictator,d2017push} and in the field \citep{frey2004social,croson2010gendered,cialdini1991focus,ferraro2013using,agerstrom2016using,goldstein2008room,hallsworth2017behavioralist}.}

Here, we extend this line of literature by studying the effect of moral suasion in dictator games in which dictators can favour a member of their own group, at a cost to a member of another group. This is an important case for two reasons. 

One is practical. Scholars have been studying ingroup favouritism (and its companion, outgroup derogation) for decades, with the underlying motivation that this is what ultimately generates some of the lightest and, at the same time, some of darkest expressions of human nature, democracy and civil rights, on the one hand, genocides and ethnic cleansings, on the other hand \citep{tajfel1971social,wilson1975theory,nowak2006five,puurtinen2008between,rusch2014evolutionary}. Therefore, exploring whether and how moral suasion affects ingroup favouritism may have significant practical implications.

The second reason is that theoretical predictions are, a priori, not obvious, as we argue below. 

On the one hand, ingroup favouritism is considered to be one of the most fundamental behavioural bias among humans \citep{haidt2012righteous,baron2013duty,curry2019good}. The seminal work by Tajfel \emph{et al}. (\citeyear{tajfel1971social}) found that people discriminate between groups even when groups are assorted through a random, irrelevant categorization (see also \cite{tajfel1970experiments}; \cite{tajfel1974social}; \cite{tajfel1982social}). Since then, ingroup favouritism has been observed in several economic contexts, ranging from public goods games \citep{krupp2008cue}, dictator games \citep{whitt2007dictator,chen2009group}, charitable donation games \citep{pavey2011highlighting}, ultimatum games \citep{mcleish2011social,kubota2013price}, prisoner’s dilemmas \citep{ahmed2007group}, and response games \citep{chen2009group}, among many others \citep{everett2015preferences}. Ingroup favouritism is so widespread that some psychologists have come to suggest that its psychological basis, \emph{group identity}, ultimately descends from the uniquely human awareness of the inevitability of death: when people identify with a group, they embed themselves into something greater, that ultimately outlives the individual, and this allows the individual to reach the so-called ``symbolic immortality'' \citep{becker2007denial,harmon1996effects,arndt1997suppression}. Important for our work is that ingroup favouritism is widespread also as a moral rule: ``help your group'' has been recently found to be a universal moral rule across 60 societies \citep{curry2019good}. This is consistent with the morality-as-cooperation theory \citep{curry2016morality}, as well as with moral foundations theory, according to which ingroup favouritism represents one of the fundamental dimensions of morality \citep{graham2013moral,haidt2012righteous,haidt2004intuitive}.

Therefore, this line of literature suggests that moral suasion might increase ingroup favouritism, by making it salient that the right thing to do is to help your group.

Alternatively, it is possible that the effect of moral suasion on ingroup favouritism is not domain-general, but it depends on specific factors such as the decision context and the individual characteristics of the decision maker. In this paper, we focus on one contextual factor and four individual characteristics.

Among the contextual factors, the way the groups are assorted is likely to affect the strength of ingroup favouritism and, ultimately, the effect of moral suasion. For example, if people are grouped according to characteristics that are central to their identity, it is likely that they would display more ingroup bias, compared to situations in which they are grouped according to less central characteristics; this might have the effect that moral suasion increases ingroup favouritism in cases in which participants are assorted according to characteristics that generate a strong group identity, while leaving it unaffected, or even decreasing it, when participants are assorted according to characteristics that generate a weaker group identity. What characteristics are central to people’s identity? Previous work has shown that people consider their moral traits to be the most essential part of their identity, even more so than emotional and autobiographical memory \citep{strohminger2014essential}. Additionally, morality is the first characteristic that people use to form impressions and evaluate others \citep{goodwin2014moral,wojciszke1998dominance}. This suggests that it is possible that assorting participants according to their moral preferences would generate a stronger group identity compared to when participants are assorted according to non-moral preferences; and this difference in group identity would translate into a difference in ingroup favouritism and the effect of moral suasion on ingroup favouritism.

Coming to the individual characteristics, there are many that, in principle, might play a role. To select some, we took a pragmatic approach. Since we planned to conduct our experiment on Prolific and since Prolific allows experimenters to download some individual characteristics of the participants, we checked, prior to the experiment, the list of all the individual characteristics that were available for download and we selected those that we believed to be potentially relevant. In doing so, we selected four characteristics: gender, political orientation, religiosity, and pro-life vs pro-choice ethical convictions. We chose gender because previous research has found that males tend to display greater levels of social dominance orientation than females do \citep{sidanius1994social}, and social dominance orientation is known to be predictive of various forms of outgroup derogation, including racism \citep{sidanius1993racism}. A related line of work suggests that between-groups competition might be evolved especially among men, because it allowed men to gain access to mates and other resources \citep{vugt2007gender,mcdonald2012evolution}. Therefore, it is possible that men display more ingroup favouritism and this could be even strengthened by moral suasion. We chose political orientation because previous work suggests the existence of a “prejudice gap” between conservatives and liberals \citep{chambers2013ideology}, with conservatives being more intolerant towards outgroups than liberals (see \citealp{sibley2008personality}, for a meta-analysis). Consequently, it is possible that moral suasion decreases ingroup favouritism for liberals, while increasing it for conservatives. For similar reasons, we chose religiosity and pro-life vs pro-choice ethical convictions, being typically correlated with political conservatism \citep{malka2012association}. See also \cite{enke2019moral}.

Following this line of thoughts, we designed, pre-registered, and conducted a well-powered (N=502), 2$\times$2 mixed-design experiment, in which the first, between-subjects, factor represents the way the groups are assorted (according to moral vs non-moral preferences), while the second, within-subject, factor represents the way people are asked to make dictator game decisions between individuals in their own group and individuals in the other group (baseline vs under moral suasion). We chose to use a within subject design for the dictator game decisions in order to be able to categorize participants in three types: those who, in response to moral suasion, discriminate more between groups; those who, in response to moral suasion, discriminate less between groups; those who, in response to moral suasion, do not change their strategy. This subdivision in types can help us shed light on the heterogeneity on people's moral preferences.\footnote{Implementing a within-subject design comes also with some costs. On the one hand, the effect of moral suasion might decrease, because a within-subject design introduces demand for consistency \citep{samuelson1988status}; on the other hand, the effect of moral suasion might increase, due to experimenter demand effect. However, we believe this last issue to be less relevant in our case, because ``demand effects refer to changes in behaviour due to cues about what constitutes appropriate behaviour'' \citep{zizzo2010experimenter}; therefore, people who change donation because of demand effect are still following a norm.}

In a nutshell, our main results are: (i) participants tend to favour their own group to a greater extent when groups are assorted according to moral preferences, compared to when they are assorted according to non-moral preferences; (ii) the net effect of moral suasion is to decrease ingroup favouritism, but there is also a non-negligible proportion of participants for whom moral suasion increases ingroup favouritism; (iii) the effect of moral suasion is substantially stable across group assorting and the four individual characteristics under consideration.

\section{Method} 
We conducted an online 2$\times$2 experiment on Prolific \citep{palan2018prolific}, implemented in oTree \citep{ chen2016otree}. We recruited 502 participants living in the US at the time of the experiment. In the first stage of the experiment, participants were randomly assigned to one of two treatments designed to group together individuals with similar stated preferences. In one treatment, preferences are collected as answers to questions on issues that are morally relevant; in the other treatment, preferences are collected as answers to questions on issues that are not morally relevant. In the second stage, each participant played a randomized sequence of three variants of the dictator game (DG) in two distinct settings: the baseline setting and the moral suasion setting. The DG is a non-strategic game, where the decision-maker, the \emph{dictator}, has to decide how to split a certain amount of money between herself and the \emph{receiver}. In our study, the decision maker has to decide how to split 100 points (in 10-points increments) between a given pair of recipients. In the next subsections, we describe the experiment in more details. Full experimental instructions are reported in Appendix A.

\subsection{Stage 1: Group Formation}
In the first stage of the experiment, participants were randomly assigned to one of the two treatments: the ``moral assorting'' (250 participants) and the ``non-moral assorting'' (252 participants) treatments. In the moral assorting treatment, participants answered five questions concerning moral issues (see Appendix A) and had to indicate if they believe these issues are morally acceptable, morally wrong or if they have no opinion. In the non-moral assorting treatment, participants were asked their preferences on five non-moral issues (see Appendix A). 
For each subject, her \emph{own group} was defined as the set of participants who answered in the same way as she did to at least three out of the five questions, with the remaining participants forming the \emph{other group}.

\subsection{Stage 2: Dictator Games}
In the second stage of the experiment, participants played DGs in two distinct settings with fixed order: first in the ``baseline'' setting and then in the ``moral suasion'' setting. In the baseline, each participant had to decide how to split 100 points in three different randomized DGs. In the first DG (which we call ``DG own''), each participant had to divide points between herself and a randomly picked member of her own group. 
In the second DG (which we call ``DG other''), each participant had to divide points between herself and a randomly picked member of the other group. In the third DG (which we call ``DG own-other''), each participant had to divide points between a randomly picked member of her own group and a randomly picked member of the other group. After this, participants played in the moral suasion setting. Here, participants faced the same three DGs described above, but before allocating points they were told: ``do what you think is morally right''. 

Lastly, participants were asked comprehension questions. We refer to Appendix A for full experimental instructions. 

After collecting all the data, participants were randomly assigned to the role of decision-maker or receiver. Then one of the DGs was randomly selected and subjects were paired according to the selected DG. Participants received a payment according to their role in that particular DG. This payment methodology implies that all decisions have an impact on both the decision maker and the recipient. On average, participants gained 0.61 GBP, including the show-up fee (0.40 GBP).

\subsection{Measures of Ingroup Favouritism}

We operationalize ingroup favouritism through two different individual-level measures. One measure is constructed starting from the DGs in which the decision-maker is affected by her decision (the ``DG own'' and the ``DG other''). Specifically, this measure is computed as the difference between how much a dictator gives to a randomly picked member of her own group and how much she gives to a randomly picked member of the other group. We call this measure \emph{costly ingroup favouritism}. Note that this is different from the notion of parochial altruism, which requires the action to be (i) costly to the decision-maker, (ii) beneficial for the decision maker's ingroup, and (iii) costly for outgroup members, all at the same time \citep{bohm2018psychology,choi2007coevolution}. In particular, costly ingroup favouritism does not necessarily involve harming outgroup members; it might be the case that a dictator still gives some amount to outgroup members, but less compared to ingroup members. We also consider a measure of \emph{costless ingroup favouritism}, whereby helping one's own group does not cost anything to the decision maker. This measure is constructed starting from the DG in which the dictator is not affected by her decision (``DG own-other''). Specifically, this measure is computed as the difference between 50 (the equal split) and how much a dictator gives to a randomly picked member of the other group. 

\subsection{Research questions}

We pre-registered three research questions:
\begin{enumerate}
    \item Does assorting based on moral preferences generate more ingroup favouritism than assorting based on non-moral preferences?
    \item Does moral suasion mitigate ingroup favouritism?
    \item Does moral suasion affect ingroup favouritism differently when group assorting is based on moral preferences compared to when it is based on non-moral preferences?
\end{enumerate}

Furthermore, we pre-registered that we would test the role of: gender, political orientation, religiosity, pro-life vs pro-choice ethical convictions. The pre-registration is available at: https://aspredicted.org/k4r34.pdf.

\section{Results}
\begin{figure}[h!]
\includegraphics[width=\columnwidth]{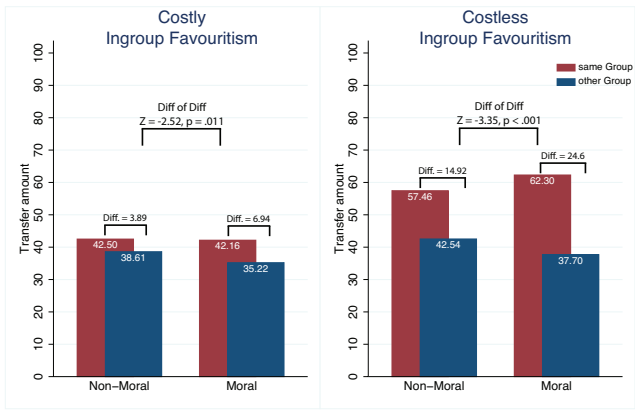}
\caption{\textbf{Ingroup favouritism is stronger when assorting is based on moral preferences, compared to when it is based on non-moral preferences.} The average of the costly ingroup favouritism measure in the moral treatment (M=6.94, SD=12.27) is 3.05 points points higher than it is in the non-moral one (M=3.89, SD=16.57)  (left chart).  The average of the costless ingroup favouritism measure in the moral treatment (M=12.3, SD=20.37) is 4.83 points higher  than it is in the non-moral one (M=7.46, SD= 15.69)  (right chart).}
\label{fig1}
\end{figure}

Our first research question is whether assorting based on moral preferences generates more ingroup favouritism compared to the case where assorting is based on non-moral preferences. Figure \ref{fig1} suggests that, on average, the answer is positive for both measures of ingroup favouritism. Wilcoxon rank-sum test confirms this finding. When assorting is based on moral preferences, the average of the costly ingroup favouritism measure is 3.05 points greater than it is when assorting is based on non-moral preferences, Z=-2.52, p=.011 (left chart). A similar result holds for the costless ingroup favouritism measure. In the moral treatment, the average of the costly ingroup favouritism measure is 4.83 points greater than it is in the non-moral treatment, Z=-3.35, p<.001 (right chart).

\begin{figure}[h!]
\includegraphics[width=\columnwidth]{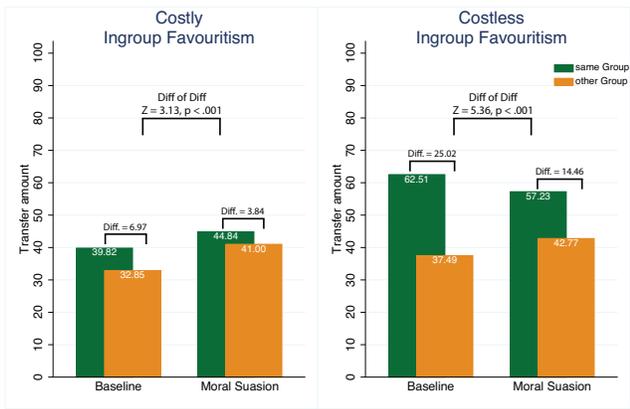}
\caption{\textbf{Ingroup favouritism is weaker under moral suasion compared to the baseline.} The average of the costly ingroup favouritism measure under moral suasion (M=3.84, 12.95)  is 3.13 points lower than it is in the baseline (M=6.97, SD= 16.02) (left chart). The average of the costless ingroup favouritism measure under moral suasion (M=7.23, SD=16.22) is 5.28 points lower than it is in the baseline (M=12.51, SD=19.89) (right chart).}
\label{fig2}
\end{figure}

The second research question is whether moral suasion mitigates ingroup favouritism compared to the baseline. Figure \ref{fig2} suggests that, on average, the answer is positive for both measures of ingroup favouritism. Wilcoxon rank-sum test confirms this finding. In the baseline, the average of the costly ingroup favouritism measure is 3.13 points higher than it is under moral suasion Z=3.13, p<.001 (left chart). A similar result holds for the costless ingroup favouritism measure. In the baseline, the average of the costless ingroup favouritism measure is 5.28 points higher than it is under moral suasion Z(502)=5.37, p<.001 (right chart). Splitting the effect of moral suasion by DG decision, we find that moral suasion increases prosociality both in ``DG own'' and ``DG other'' (linear regression: $t=4.47$, $p<.001$; $t=4.61$, $p<.001$). However, the increase of prosociality in ``DG other'' is even greater ($t=6.40$, $p<.001$), and this is ultimately the reason why, on average, moral suasion attenuates ingroup favouritism.

The fact that, on average, moral suasion decreases ingroup favouritism does not imply that moral suasion has the effect of decreasing ingroup favouritism for all the participants. In fact, we now use the within-subject factor to show that there are three non-negligible types of participants, which characteristically differ on how they react to moral suasion. In the first type we include participants who increased their ingroup favouritism under moral suasion (\textit{persuaded parochialists}: ingroup favouritism under moral suasion is higher than ingroup favouritism in the baseline\footnote{In this case we use the word \emph{parochialism}, because persuaded parochialists actually act in a way that is costly to outgroup members, compared to the baseline. Therefore, all three properties listed by \cite{bohm2018psychology} that classify parochial decisions are satisfied.}). In the second type we classify participants who decrease their ingroup favouritism under moral suasion (\textit{persuaded universalists}: ingroup favouritism under moral suasion is lower than ingroup favouritism in the baseline). In the third type we categorize  participants who do not change behaviour under moral suasion (\textit{unpersuaded}: ingroup favouritism under moral suasion is equal to ingroup favouritism in the baseline). Table \ref{tab: table1} reports the proportions of these different behavioural types in each treatment and using both the costly and costless ingroup favouritism measures. Across treatments and measures, the majority of participants is unpersuaded (on average, 72.2\%), a substantial proportion is persuaded universalist (on average, 20.5\%), and, interestingly, a smaller but non-negligible  proportion is persuaded parochialist (on average, 7.3\%). To provide further evidence that persuaded parochialist are not driven by error, we distinguish persuaded parochialists in the costly setting from those of the costless setting. In particular, we construct two variables: one takes value 1 if a participant is classified as persuaded parochialist according to the costly ingroup favouritism measure, and 0 otherwise; and the other is analgously defined by using the costless ingroup favouritism measure. Pearson's Chi-squared test shows that these two variables are not independent ($p<.001$). Since these measures have been collected separately, this suggests that they are sensitive to the same causal factor.

\begin{table}[h!]\centering
\caption{Frequencies of the different types using costly and costless ingroup favouritism measures, across treatments.}
\label{tab: table1}
\resizebox{\columnwidth}{!}{%
\begin{tabular}{l*{4}{c}}
\toprule
            &\multicolumn{2}{c}{Costly}&\multicolumn{2}{c}{Costless}\\
            \cmidrule(lr){2-3}\cmidrule(lr){4-5} \\
  Treatment    &\multicolumn{1}{c}{\textit{Non-Moral}}&\multicolumn{1}{c}{\textit{Moral}}&\multicolumn{1}{c}{\textit{Non-Moral}}&\multicolumn{1}{c}{\textit{Moral}}\\
\midrule
Persuaded Parochialists &        8.3\%&        8.4\%&        6.3\%&        6.0\%\\
\addlinespace
Persuaded Universalists &       19.8\%&       16.0\%&       24.6\%&       21.6\%\\
\addlinespace
Unpersuaded &       71.8\%&       75.6\%&       69.1\%&       72.4\%\\
\bottomrule
\mc{5}{p{3.5in}}{Note. \textit{Persuaded parochialists}: participants who increase ingroup favouritism under moral suasion, compared to the baseline. \textit{Persuaded universalists}: participants who decrease ingroup favouritism under moral suasion, compared to the baseline. \textit{Unpersuaded}: participants who do not change ingroup favouritism under moral suasion, compared to the baseline.}
\end{tabular}
}
\end{table}

\begin{figure}[h!]
\includegraphics[width=\columnwidth]{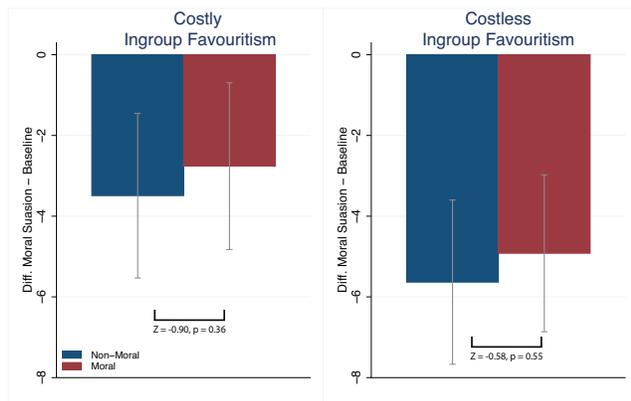}
\caption{\textbf{The effect of moral suasion on ingroup favouritism is substantially stable across group assorting.} The reduction in the costly ingroup favouritism measure due to moral suasion in the moral treatment is not significantly different from its reduction in the non-moral treatment (left chart, p-value= .36). The reduction in the costless ingroup favouritism measure due to moral suasion in the moral treatment is not significantly different from its reduction in the non-moral treatment (right chart, p-value= .55).  Error bars represent 95\% CI.}
\label{fig3}
\end{figure}

We now move to the third research question, whether there is any difference in how moral suasion affects ingroup favouritism in the two treatments, i.e., when assorting is based on moral preferences and when it is based on non-moral ones. Figure \ref{fig3} reports the difference in ingroup favouritism under moral suasion and the baseline, across treatments and measures. Mixed-design ANOVA predicting ingroup favouritism as a function of group assorting, moral suasion, and their interaction, shows no significant interaction, both for the costly ingroup favouritism measure ($F=0.25, p=0.620$) and for the costless one ($F=0.25, p=0.617$). This result is confirmed if we use the rank-sum test. Using both measures of ingroup favouritism, we find that the effect of moral suasion is substantially stable across group assorting (costly measure: Z=-.90, p=.36; costless measure: Z=-.59, p=.55). 
Table \ref{tab: table2} reports the fractions of participants for whom the ingroup favouritism measure is strictly greater than 0, across measures and treatments. Again we find that the effect of moral suasion is substantially stable across treatments, as confirmed by mixed-design ANOVA, which reports a non-significant interaction between group assorting and moral suasion (costly measure: F=1.96, p= 0.1618; costless measure: F=0.95, p= 0.3297).


\begin{table*}[h!]\centering
\caption{Percentages of participants for which the ingroup favouritism measure is strictly greater than 0, across treatments and ingroup favouritism measures.}
\label{tab: table2}
\begin{tabular}{L{.9in}C{.9in}C{.9in}C{.9in}C{.9in}C{.9in}C{.9in}}\toprule
 & \mc{3}{c}{ Costly} & \mc{3}{c}{Costless} \\
\cmidrule(lr){2-4}\cmidrule(lr){5-7} \\
 Treatment & \itshape Baseline & \itshape Moral Suasion &  Difference & \itshape Baseline & \itshape Moral Suasion &
 Difference\\ \midrule
\textit{Non-Moral}	& 24.9\% & 11.3\% & 13.6\% & 36.7\% & 17.5\% & 19.2\% \\
 \textit{Moral}	 & 27.7\% & 20.0\% & 7.7\% & 43.5\% & 28.6\% & 14.9\%  \\
\bottomrule
\end{tabular}
\end{table*}

Finally, we ask whether the differential effect of moral suasion depends on the four pre-registered individual characteristics that we could download from Prolific: gender, political orientation, religiosity, and pro-life vs pro-choice  ethical convictions. To do so we coded 4 dummy variables: \textit{gender}, 1 if female and 0 if male; \textit{liberal}, 1 if Liberal and 0 if Conservative/Moderate; \textit{no\_religious}, 1 if agnostic, atheist or no religious, and 0 if religious; \textit{prochoice}, 1 if a participant reports to have pro-choice ethical convictions and 0 if a participant reports to have pro-life ethical convictions. Table \ref{tab: table3} reports average characteristics of participants by treatments and types. The type subdivision is substantially stable across treatments and individual characteristics. Females, liberals, no religious and pro-choice people tend to be slightly more represented among persuaded universalists than they are among persuaded parochialists. However, the only statistically significant difference is that persuaded universalists are less religious than persuaded parochialists, Z(61)=-2.69, p=.007. Therefore, we conclude that the effect of moral suasion is substantially stable across the four individual characteristics that we considered.

\begin{table}[h!]\centering
\caption{Average of different groups' characteristics in the non-moral and moral treatments.}
\label{tab: table3}
\resizebox{\columnwidth}{!}{%
\begin{tabular}{l*{6}{c}}
\toprule
&\multicolumn{2}{c}{Persuaded parochialist}&\multicolumn{2}{c}{Persuaded universalist}&\multicolumn{2}{c}{Unpersuaded}\\
\cmidrule(lr){2-3}\cmidrule(lr){4-5}\cmidrule(lr){6-7} \\
Treatment &\multicolumn{1}{c}{\textit{Non-Moral}}&\multicolumn{1}{c}{\textit{Moral}} &\multicolumn{1}{c}{\textit{No-Moral}} &\multicolumn{1}{c}{\textit{Moral}}&\multicolumn{1}{c}{\textit{Non-Moral}} &\multicolumn{1}{c}{\textit{Moral}}\\
\midrule
Gender      &        0.55&        0.52&        0.60&        0.65&        0.59&        0.53\\
Liberal     &        0.53&        0.37&        0.43&        0.58&        0.58&        0.59\\
No Religious    &        0.62&        0.43&        0.70&        0.78&        0.65&        0.62 \\
Pro Choice   &        0.83&        0.76&        0.84&        0.86&        0.82&        0.75\\
\bottomrule
\mc{7}{p{4.5in}}{Note. \textit{Gender}: 1 if female; 0 if male. \textit{Liberal}: 1 if Liberal; 0 if Conservative/Moderate. \textit{No religious}:1 if agnostic, atheist or no religion; 0 if religious. \textit{Pro Choice}: 1 if pro-choice ethical convictions; 0 if pro-life ethical convictions.}
\end{tabular}
}
\end{table}

We also tested the effect of the four individual characteristics on baseline levels of ingroup favouritism. For each individual characteristics, we used mixed-design ANOVA predicting ingroup favouritism as a function of assorting, moral suasion, the individual characteristics, and all their two- and three-way interactions. The three-way interaction is significant only in the case of \emph{liberal} and only for the costly ingroup favouritism measure ($F=3.0$8, $p = 0.04$). However, the interaction between being liberal and group assorting is strongly significant for both measures (costly: $F=8.12$, $p=0.004$; costless: $F=7.11$, $p=0.007$). The positive coefficient suggests that ingroup favouritism is maximum among liberals assorted according to their moral preferences. Regarding the effect of religiosity, we find that it marginally significantly interacts with moral suasion (costly: $F=3.37$, $p=0.067$; costless: $F=3.53$, $p=0.061$), and it significantly interacts with moral assorting, but only in the case of the costless measure ($F=4.10$, $p=0.043$). Regarding the effect of pro-life vs pro-choice ethical convictions, we found only a marginally significant interaction with group assorting and only for the costly measure ($F=3.34$, $p=0.068$). Finally, regarding the effect of gender, we found no significant interactions (all $p$'s $>$ 0.1).

\section{Discussion}
The interest in moral suasion stems, at least in part, from being a cheap and possibly effective policy tool that could be applied to foster prosocial behaviours. While the literature on moral behaviour has so far produced a substantial body of evidence showing the effectiveness of moral suasion, its dependence on the identity of the recipients of the decision-maker's actions is far less studied, leaving open the possibility that individuals react to moral suasion by reducing prosociality towards some \emph{types} of recipients. This paper has addressed this issue in the setting of a decision to split a given amount of money between members of one's own group and members of another group, providing experimental evidence that, \emph{on average}, moral suasion increases pro-sociality towards both the ingroup and the outgroup; however, the increase towards the outgroup is greater than the increase towards the ingroup, and this results in the fact that ingroup favouritism, on average, declines under moral suasion. This effect exists when groups are defined in terms of similarity with respect to answers to both non-morally relevant questions and morally relevant questions, even if, in the latter case, the initial level of ingroup favouritism is higher. We would like to stress that we are not suggesting that moral assorting is \emph{qualitatively} different from non-moral assorting. The fact that moral assorting generates stronger initial ingroup favouritism compared to non-moral assorting is likely to be driven by the fact that moral assorting makes group identity \emph{quantitatively} stronger. What is important to note, instead, is that the negative effect of moral suasion on ingroup favouritism holds \emph{on average}. When we look at how participants change their decision in response to moral suasion, we find that there is a small, but non-negligible, proportion of participants who increase their level of ingroup favouritism. Finally, the effect of moral suasion appears to be substantially stable across four (pre-registered) individual characteristics: gender, political orientation, religiosity, pro-choice vs pro-life ethical convictions. The only significant effect is that persuaded universalists tend to be less religious than persuaded parochialists. These findings have potential applications outside the laboratory, as they suggest that making the morality of an action salient might be a practical and effective tool for decreasing ingroup favouritism, on average, while also having the drawback of actually increasing ingroup favouritism for a non-negligible proportion of participants. 

This study also relates to research exploring whether moral assorting affects ingroup favouritism. \cite{parker2013lessons} showed that when people are assorted according to their preferences on abortion, then they report feeling stronger positive ingroup emotions and negative outgroup emotions, compared to when they are assorted according to whether they prefer the Red Sox or the Yankees (this study was conducted in Massachussets, were the rivalry between these two teams is particularly strong). \cite{weisel2015ingroup} divided people in groups according to their preferences about football clubs vs political parties and found that people actively harm outgroup members only in the case in which assorting is based on political preferences. Our work is conceptually in line with this literature as our first result shows that people display greater ingroup favouritism when they are assorted according to moral preferences, compared to when they are assorted according to non-moral preferences.

This is the first study investigating the effect of moral suasion on ingroup favouritism. As such, it does have several limitations that might suggest directions for future research. First, we focused on only one contextual factor that might impact the effect of moral suasion on ingroup favouritism: assorting according to moral preferences. Future work should explore how moral suasion affects ingroup favouritism when group assorting is based on other characteristics that are likely to activate a strong group identity. For example, scholars agree that ingroup favouritism is rooted in our evolutionary tribal past  \citep{wilson1975theory,nowak2006five,puurtinen2008between,rusch2014evolutionary,fu2012evolution,masuda2015evolutionary}. This suggests that characteristics that provide group advantages (e.g., language) might be better candidates than characteristics that primarily provide individual advantages (e.g., skill specialization). Such an investigation could lead to identify specific cases where moral suasion is particularly effective, and possibly others where it delivers undesirable effects. Another potential variable of interest is the mode of cognition -- whether decisions are made under ``system 1'' (often referred to as \emph{intuition}) or ``system 2'' (often referred to as \emph{deliberation}) \citep{evans2013dual}. One study found that promoting intuition favours ingroup favouritism \citep{de2015intergroup}; two more found that promoting intuition favours cooperation with outgroup members \citep{rand2015social,everett2017deliberation}. See \cite{capraro2019dual} for a review. Consequently, it is possible that the effect of moral suasion on ingroup favouritism is moderated, or even mediated, by the mode of cognition. Future work should explore this possibility. A similar limitation regards individual characteristics. In this work we focused on gender, political orientation, religiosity, and pro-life vs pro-choice ethical convictions. The results suggest that the effect of moral suasion is substantially stable across these characteristics. Future work should explore the role of other personal characteristics. Another limitation regards the fact that our results are based on a laboratory experiment. Since moral suasion potentially represents a very practical tool for policy interventions, a key direction for future work is to test the external validity of our findings.

In sum, we studied the effect of moral suasion on ingroup favouritism. Our main results are: (i) in the baseline, participants tend to favour their own group to a greater extent when groups are assorted according to moral preferences, compared to when they are assorted according to non-moral preferences; (ii) the net effect of moral suasion is to decrease ingroup favouritism, but there is also a non-negligible proportion of participants for whom moral suasion increases ingroup favouritism; (iii) the effect of moral suasion is substantially stable across group assorting and four pre-registered individual characteristics (gender, political orientation, religiosity, pro-life vs pro-choice ethical convictions). Future work should test the effect of other contextual factors and individual characteristics.


\bibliographystyle{apa}

\bibliography{main}



\bigskip
\section*{Appendix: Experimental instructions} 
\subsection*{Group Formation}
\textit{Participants were randomly divided into two treatments. We report the instructions for both treatments}.\\
\\
\textbf{Assorting based on Moral preferences:}\\
Below you see a list of issues. For each one of them, regardless of whether or not you think it should be legal, please indicate whether you personally believe that in general it is morally acceptable, morally wrong or if you have no opinion:
\begin{itemize}
\item Abortion: Morally Acceptable/Morally Wrong/No Opinion;
\item Doctor assisted suicide: Morally Acceptable/Morally Wrong/No Opinion;
\item Death penalty: Morally Acceptable/Morally Wrong/No Opinion;
\item Gay or lesbian relations: Morally Acceptable/Morally Wrong/No Opinion;
\item Prostitution: Morally Acceptable/Morally Wrong/No Opinion.
\end{itemize}
\textbf{Assorting based on Non-Moral preferences:}\\
Below you see a list of questions. For each one of them, please indicate which option you personally prefer, or if you have no opinion.
\begin{itemize}
\item Where do you prefer to go during vacation? Sea/Mountain/No Opinion;
\item Where do you prefer to watch movies? Movie Theater/Home/No Opinion;
\item Where do you prefer to do physical activity? Gym/Outdoor/No Opinion;
\item Which social network do you prefer? Instagram/Facebook/No Opinion;
\item Which animal do you prefer? Dog/Cat/No Opinion.
\end{itemize}
\textit{Here participants moved to the next screen.}\\
Two groups will be formed, YOUR GROUP and the OTHER GROUP, using the answers that you and the other participants have given so far. YOUR GROUP is formed by you and by those participants with answers most similar to yours. OTHER GROUP is formed by the remaining participants, those with answers least similar to yours.

\subsection*{Ingroup Favouritism}
\textit{Here participants faced three randomized DG in the baseline setting.}\\
In the next screens, you will make a number of decisions about how to divide 100 Points between two participants, drawn from either YOUR GROUP, the OTHER GROUP or both. Once the survey is over, payments will be determined according to either one of the decisions you made or one of the decisions made by another participant that involves you. 100 Points correspond to payment of 0.50 GBP.
\begin{itemize}
\item You have to allocate Points between YOU and another member of YOUR GROUP.
\item You have to allocate Points between YOU and a member of the OTHER GROUP.
\item You have to allocate Points between a member of YOUR GROUP (not you) and a member of the OTHER GROUP.
\end{itemize}
\textit{Here all participants faced three randomized DG in the moral suasion setting.}\\
In the next decisions, you have to decide how to divide Points according to what you think is morally right.
\begin{itemize}
\item You have to allocate Points between YOU and another member of YOUR GROUP.
Do what you think is morally right.
\item You have to allocate Points between YOU and a member of the OTHER GROUP.
Do what you think is morally right.
\item You have to allocate Points between a member of YOUR GROUP (not you) and a member of the OTHER GROUP.
Do what you think is morally right.
\end{itemize}
\textit{Here the comprehension questions}.
\begin{itemize}
\item What is the decision that lets you obtain the highest payment?
\item What is the decision that lets the other participant obtain the highest payment?
\item What is the decision that lets you and the other participant obtain the same payment?
\end{itemize}

\end{document}